\documentclass[]{aa}

\usepackage{txfonts}
\usepackage{graphicx}

\begin{document}

\title{A Corona Australis cloud filament seen in NIR scattered light
\thanks{Based on observations made with ESO telescopes at the La Silla
Paranal Observatory under programme ID 077.C-0338.}
}
\subtitle{I. Comparison with extinction of background stars}

\author{
Mika Juvela \inst{1} \and
Veli-Matti Pelkonen \inst{1} \and
Paolo Padoan \inst{2} \and
Kalevi Mattila \inst{1} 
}

\institute{
Helsinki University Observatory, FIN-00014, University of Helsinki, Finland
\and
Department of Physics, University of California, San Diego, 
CASS/UCSD 0424, 9500 Gilman Drive, La Jolla, CA 92093-0424; ppadoan@ucsd.edu
}

\date{Received 1 January 2005 / Accepted 2 January 2005}

\abstract
% context
{With current near-infrared (NIR) instruments the near-infrared light 
scattered from interstellar clouds can be mapped over large areas. The surface
brightness carries information on the line-of-sight dust column density.
Therefore, scattered light could provide an important tool to study mass
distribution in quiescent interstellar clouds at a high, even sub-arcsecond
resolution. }
% aims
{We wish to confirm the assumption that light scattering dominates the
surface brightness in all NIR bands. Furthermore, we want to show
that scattered light can be used for an accurate estimation of dust column
densities in clouds with $A_{\rm V}$ in the range 1-15$^{\rm m}$. }
% methods
{We have obtained NIR images of a quiescent filament in the Corona Australis
molecular cloud. The observations provide maps of diffuse surface brightness
in J, H, and Ks bands. Using the assumption that signal is caused by scattered
light we convert surface brightness data into a map of dust column
density. The same observations provide colour excesses for a large number of
background stars. These data are used to derive an extinction map of the
cloud. The two, largely independent tracers of the cloud structure are compared. }
% results
{In regions below $A_{\rm V}\sim 15^{\rm m}$ both diffuse surface brightness
and background stars lead to similar column density estimates. The existing
differences can be explained as a result of normal observational errors and
bias in the sampling of extinctions provided by the background stars.  There
is no indication that thermal dust emission would have a significant
contribution even in the Ks band. The results show that, below $A_{\rm V}\sim
15^{\rm mag}$, scattered light does provide a reliable way to map cloud
structure. Compared with the use of background stars it can also in practice
provide a significantly higher spatial resolution.
}
% conclusions
{}


\maketitle

\section{Introduction}

Distribution of matter within interstellar clouds can be examined only
indirectly by measuring emission or absorption of radiation. Extended maps of
interstellar clouds are obtained through (1) emission lines of molecules or
atomic tracers, especially CO and HI, (2) thermal emission from dust grains at
far--IR and sub-mm wavelengths, (3) optical or near-infrared extinction as
traced by star counts, (4) optical or near-infrared reddening seen in the
light of background stars, (5) mid-infrared absorption in dark clouds seen
against a brighter background, and (6) soft x-ray absorption toward a
similarly bright x-ray background. 

The correspondence between measured quantities and the actual column densities
is not straightforward. Line emission is useful only in a limited range,
defined by the chemistry, critical density, and optical depth. The
interpretation of measured intensities is further complicated by spatially
varying excitation conditions and other radiative transfer effects.  Emission
from different molecules is often observed to peak at entirely different
locations and our view of a cloud's structure can be seriously biased by the
selection of certain tracers. In particular, in cold cloud cores most
molecules may have frozen onto dust grains, making them effectively invisible in
studies of many molecular lines. At low column densities ($A_{\rm V}\sim$1)
the transition between atomic to molecular gas poses similar problems because
of the large abundance variations.

Thermal dust emission at far-IR and sub-mm wavelengths provides a
complementary tool which, however, is also not free of problems. Conversion
into column density requires reliable estimates of the dust temperature
and emissivity. Recent studies have shown, that there may be significant
variations between clouds and even locally within individual sources
(Cambr\'esy et al. \cite{cambresy2001}, del Burgo et al. \cite{delburgo2003},
Dupac et al. \cite{dupac2003}; Kramer et al. \cite{kramer2003}, Stepnik et al.
\cite{stepnik2003}; Lehtinen et al. \cite{Lehtinen2004}, \cite{Lehtinen2006};
Ridderstad et al. \cite{Ridderstad2006}). These may be due to grain growth by
coagulation and ice mantle deposition (Ossenkopf \& Henning
\cite{ossenkopf1994}, Krugel \& Siebenmorgen \cite{krugel1994}) or even
physical changes in the grain material itself (e.g., Mennella et al.
\cite{mennella98}, Boudet et al. \cite{Boudet2005}). Limited sensitivity of
observations has restricted most sub-mm emission studies to areas
of large column density and/or embedded heating sources. In those cases the
spatial temperature and emissivity variations tend to be large, and the density
structure can be estimated only indirectly, through complicated modelling.

The problems caused by temperature variations can be avoided by looking at the
light attenuation caused by dust particles. Extinction can be traced with star
counts (Wolf \cite{wolf1923}) or by examining the change in the colour of
stars seen through the cloud. NIR observations have become increasingly
important in extinction studies (e.g., Alves, Lada \& Lada \cite{alves2001};
Cambr\'esy et al. \cite{cambresy2002}) because, at these wavelengths, extinction
properties of grains are believed to be very stable and background stars can
still be observed through visually opaque clouds. In the NIR the star count
method remains useful beyond $A_{\rm V}\sim$20, provided that deep K-band
observations are available. However, at lower extinctions the colour excess
method yields better spatial resolution. An independent extinction estimate is
obtained for each detected background star, i.e., a very narrow sight line.
Errors are usually dominated by the uncertainty of the intrinsic colours that
are known only in a statistical sense. An accurate extinction map is obtained
only after averaging spatially over many stars. The reliability can be
improved by combining results from more than two NIR bands (Lombardi \& Alves
\cite{Lombardi2001}) and by using adaptive spatial resolution (Cabr\'esy et al.
\cite{cambresy2002}). With dedicated observations a resolution of
$\sim$10$\arcsec$ can be reached. In the case of the commonly used 2MASS
survey (limiting K$_{\rm s}$ magnitude $\sim$15) the resolution is, depending
on the stellar density towards the examined field, a few arc minutes and the
probed range of extinctions is $A_{\rm V}\sim$1--15\,mag.

Scattered light provides yet another tracer for studies of interstellar
clouds. The first detection of NIR scattered light in a dark nebula
illuminated by the normal interstellar radiation field (ISRF) was reported by
Lehtinen and Mattila (\cite{Lehtinen1996}).  Recently Nakajima et al.
(\cite{Nakajima2003}) and Foster \& Goodman (\cite{Foster2006}) have
demonstrated that with current NIR instruments the scattered radiation can be
mapped in moderately optically thick clouds over large areas. In these
conditions there should exist a relatively simple relationship between the
observed surface brightness and the amount of dust on the line-of-sight. Based
on this idea, Padoan, Juvela, \& Pelkonen (\cite{Padoan2006}) proposed scattered
near-infrared light as a new tracer of cloud column density.  As mentioned
above, dust properties are relatively constant in the NIR, which also reduces
uncertainties related to the scattering properties.  In the K-band the optical
depth is about one tenth of the visual extinction. Therefore, in regions with
$A_{\rm V}$ below 10\,mag, the K-band intensity is expected to remain well
correlated with dust column density. At higher extinctions the surface
brightness starts to saturate. However, Padoan et al. (\cite{Padoan2006})
showed that if the saturation is taken into account, the combination of J-,
H-, and K-bands can be used for column density estimation up to $A_{\rm V}\sim
20^{\rm m}$.
Juvela et al. (\cite{Juvela2006}) examined further the relationship between
surface brightness and column density using a series of inhomogeneous cloud
models. They concluded that the errors in the column density estimates remain
small even in the presence of expected dust property variations and in the
case of anisotropic external illumination. Furthermore, they noted that
the comparison of surface brightness and colour excess methods makes it possible
to identify and correct effects caused for example by anisotropic illumination. As a
column density tracer, scattered light was estimated to be as reliable as the
use of background stars. At the same time, scattered light could provide
much higher resolution.

\begin{figure} 
\resizebox{\hsize}{!}{\includegraphics{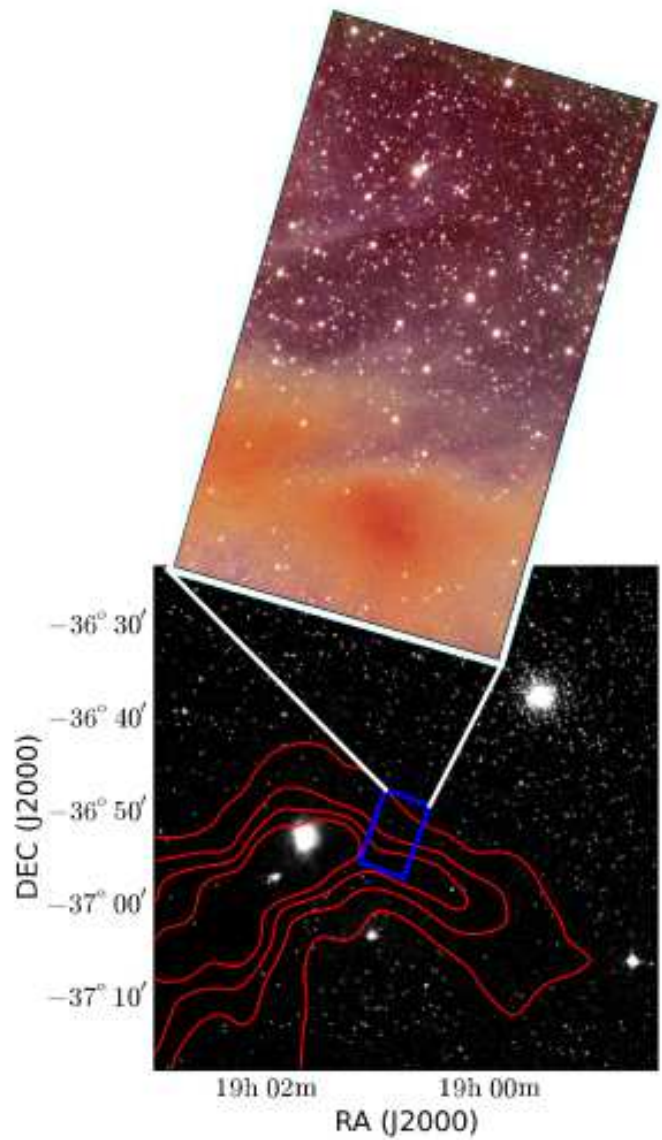}} 
\caption{
The lower frame shows a Digitized sky survey image of the northern part of the
Corona Australis Molecular cloud. The contours indicate optical extinction
determined with the NICER method using stars from the 2MASS survey. The
contours are drawn between 2 and 10 magnitudes at steps of two magnitudes. The
box shows the outlines of the area observed in the near-infrared with the SOFI
instrument. The bright region east of the mapped area is a reflection nebula
associated with the young star R Corona Australis. The upper frame shows an
image composed of our the J, H, and Ks band observations. The intensity scale
has been adjusted to bring out faint surface brightness. High J/Ks ratios
corresponds to blue, and low ratios to red colour.
} 
\label{fig:plot_dss} 
\end{figure}

In order to examine the use of scattered light as a tracer of interstellar
clouds we have carried out deep J-, H-, and Ks-band observations of a filament
in the northern part of the Corona Australis cloud.  Based on 2MASS data, at
the resolution of a couple of arc minutes, the extinction was estimated to be
in the range $A_{\rm V}=1-10\,$mag, a range suitable for this method (see
Fig.~\ref{fig:plot_dss}). The diffuse surface brightness was detected in all
three bands and, therefore, can be used for estimation of column densities
over most of the observed area. The imaging provides simultaneously photometry
for a large number of background stars so that an extinction map can be
constructed based on the colour excesses of those stars. The comparison of
surface brightness and extinction values makes it possible to estimate the
reliability of scattered light as a probe of cloud column density.
We will show that the observed NIR surface brightness is consistent with the
assumption that the signal is caused by light scattering. In particular, we
will show that below $A_{\rm V}=15^{\rm m}$ the column density estimates from
surface brightness data and from background stars are consistent with each
other.

We present in Sect.~\ref{sect:observations} the observations and in
Sect.~\ref{sect:qualitative} a qualitative analysis of the surface brightness
data. In Sect.~\ref{sect:conversion} the surface brightness measurements are
converted into a map of column density using the method outlined by Padoan et
al. (\cite{Padoan2006}). In Sect.~\ref{sect:NICER} the reddening of the light
from background stars is analyzed using the NICER method, resulting in
another, largely independent map of column density. The results are compared
with each other in Sect.~\ref{sect:comparison}. The discussion of the results
is presented in Sect.~\ref{sect:discussion} and summarized in
Sect.~\ref{sect:conclusions}. 

In this paper we limit the study to regions with $A_{\rm V}$ below
$\sim$15$^{\rm m}$ where conversion between surface brightness and column
density is more straightforward. In a forthcoming paper we will carry out
three-dimensional radiative transfer modelling of the whole field. This will
enable us to estimate or derive lower limits for column densities also in the
most optically thick regions. In that paper we will extend comparison with the
NICER method beyond $A_{\rm V}=20^{\rm m}$, and look for signs of dust
property variations within the filament.

\section{Observations} \label{sect:observations}

The field was observed with the explicit purpose of testing the conversion
between scattered light and column density. From the Corona Australis
molecular cloud we selected a region where the visual extinction was estimated
to be in the range $A_{\rm V}=1-10$ magnitudes, and where there were no strong
infrared sources within the field, either physically within the cloud or seen
in projection.  The field is centered at $19^{\rm h}0^{\rm m}51^{\rm s}$,
$-36\degr 52\arcmin 30\arcsec$ (J2000) and covers part of the filament that
extends SW from the northern end of the Corona Australis cloud. In equatorial
coordinates the filament runs horizontally across the southern end of the
image (see Fig.~\ref{fig:plot_dss}). In the northern end the image extends
outside the filament to a region of low extinction. The extinction was
originally estimated using 2MASS data and the NICER method. At the resolution
of 3 arc minutes the range of visual extinctions was $A_{\rm V}=1-10$
magnitudes (see Fig.~\ref{fig:plot_dss}), although it was also clear that
higher spatial resolution could reveal smaller areas with much higher
extinction. For the filament itself the extinction decreases towards west,
where the 2MASS extinction drops rapidly below 5 magnitudes.  The area has
been mapped in 1.2\,mm continuum by Chini et al. (\cite{Chini2003}). In those
observations the filament is not smooth but appears to have several distinct
density enhancements. This is already confirmed by Fig.~\ref{fig:plot_dss}.
Within the field imaged in NIR the brightest star has a Ks band magnitude of
12.1.

The observations were made in August 2006 using the SOFI instrument on the NTT
telescope. The observations were carried out as ON-OFF measurements so that
the faint surface brightness could be recovered. We used two ON-fields that
overlap each other by $\sim 1\arcmin$. Four OFF fields were selected based on
IRAS images and 2MASS extinction maps from regions with low dust column
density. The observations of the ON- and OFF-fields were interleaved in a
sequence of ON-2 -- OFF -- ON-1 in order to minimize the effect of sky
variations. The OFF-field was varied in different observation blocks in order
to average out any faint background gradients that might still exist in the
OFF-fields. Each observed frame was an average of six 10 second exposures,
corresponding to an integration time of 1 minute. After each sequence, the
telescope returned to ON-2. At this point random jitter was added in order to
avoid holes caused by bad pixels. The total integration times for each
ON-field was 65 minutes in J, 91 minutes in H, and 260 minutes in Ks band (see
Table~\ref{table:fields}).

\begin{table}
\caption{Positions and total integration times of the observed ON and OFF fields.
The last column gives list for the OFF fields the extinction estimates of
Schlegel, Finkbeiner \&  Davis (\cite{SFD}).}
\begin{tabular}{llllll}
Field & centre position & $t$(J) & $t$(H) & $t$(Ks)
& $A_{\rm V}$ \\
&    (J2000)      &  (min)           & (min)            & (min) &  (mag) \\
\hline
ON-1    &  $19^{\rm h}0^{\rm m}54^{\rm s}$, $-36\degr 54\arcmin 0\arcsec$
& 65  & 91  & 260 &   -- \\
ON-2    &  $19^{\rm h}0^{\rm m}48^{\rm s}$, $-36\degr 51\arcmin 0\arcsec$ 
& 65  & 91  & 260 &   -- \\
OFF-1   &  $18^{\rm h}58^{\rm m}14^{\rm s}$, $-36\degr 50\arcmin 11\arcsec$ 
& 18  & 26  & 78  &   0.45 \\
OFF-2   &  $18^{\rm h}58^{\rm m}15^{\rm s}$, $-36\degr 34\arcmin 43\arcsec$ 
& 18  & 26  & 89  &  0.35 \\
OFF-3   &  $18^{\rm h}56^{\rm m}59^{\rm s}$, $-36\degr 53\arcmin 5\arcsec$
& 18  & 26  & 80  &  0.33 \\
OFF-4   &  $18^{\rm h}57^{\rm m}34^{\rm s}$, $-36\degr 50\arcmin 11\arcsec$ 
& 11  & 13  & 13  &  0.33 \\
\end{tabular}
\label{table:fields}
\end{table}

\begin{figure} 
\resizebox{\hsize}{!}{\includegraphics{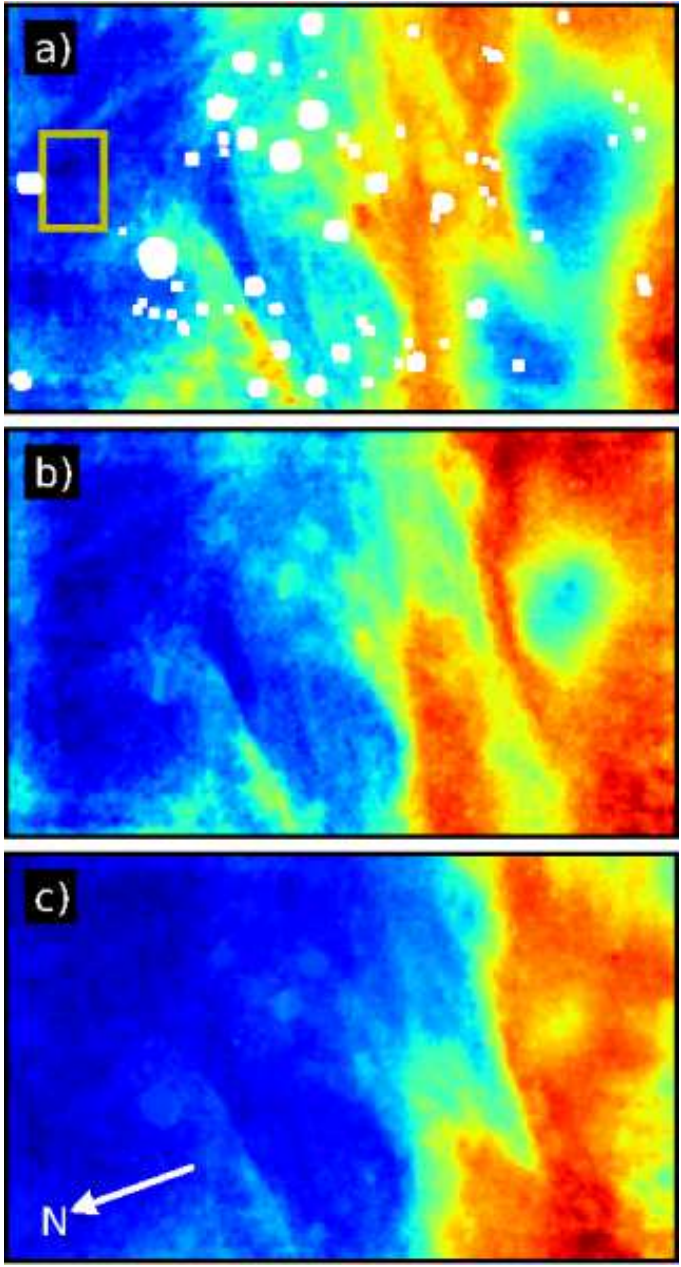}} 
\caption{
Observed surface brightness in J, H, and Ks bands (frames {\em a}, {\em b},
and {\em c}, respectively). Bright stars identified by DAOPHOT were removed
and replaced with interpolated surface brightness. Faint stars have been
eliminated with median filtering. In the J-band image (frame $a$) the white
regions correspond to areas around bright stars that were masked and removed
from subsequent analysis. The box denotes an area used as a reference point
for the NIR surface brightness.  In frame $c$ the arrow shows the direction of
equatorial north.
} 
\label{fig:JHKmaps} 
\end{figure}

The observations were calibrated to 2MASS scale by using photometry of ten
selected stars in each individual frame. Their average was used to correct the
observed fluxes to above the atmosphere. After calibration the two ON-fields
were mosaiced together using the overlapping area. This required an addition
small correction to ON-1, to have the same number of counts in the overlapping
area as ON-2. Finally, a correction was added to the combined frame to place
the minimum of the diffuse emission at zero. After removing borders with high
noise, the remaining imaged area is 8.4$\arcmin \times$5.0$\arcmin$. For
studies of the surface brightness we produced another set of images where we
try to eliminate the effect of stars. Stars were first removed using DAOPHOT
task ALLSTAR. In many cases this left noticeable residuals at the location of
bright stars. These were removed by interpolating over the masked stellar
image. This affects only the figures because in the analysis these pixels were
still kept masked.  The effect of some bright stars extends beyond the area
masked by DAOPHOT. In these cases the masks were extended manually, removing
areas were the surface brightness enhancement was visibly above the general
background. Finally, faint stars that were not identified by DAOPHOT were
removed with median filtering. The filter size was 14 pixels or 4.0 arc
seconds which is roughly equal to 4 times the FWHM of the point spread
function. The final J-, H-, and Ks-band surface brightness images are shown in
Fig.~\ref{fig:JHKmaps}. This and the subsequent images are rotated so that
equatorial north is towards lower left. In this orientation the surface
brightness increases to the right and the filament runs vertically through the
right hand side of the images. The Ks-band can be assumed to correlate best
with the column density. In J- and H-bands the optical depths are higher, the
surface brightness saturates earlier and finally decreases so that the most
opaque regions correspond to local minima. There is a small local minimum also
in Ks, at the location of presumably the highest column density. On other
sightlines, where at least Ks surface brightness is not yet completely
saturated, the scattered light should be unequivocally correlated to the
column density.

\section{Correlations between NIR surface brightness maps} \label{sect:qualitative}

\begin{figure}
\resizebox{\hsize}{!}{\includegraphics{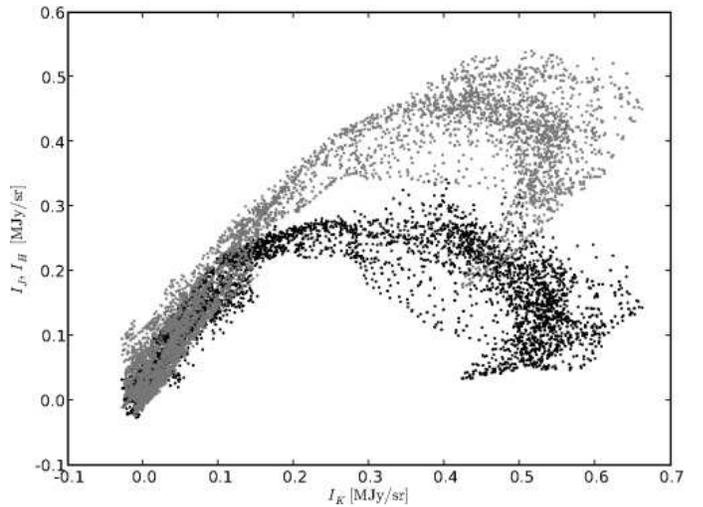}}
\caption{
The observed J band intensities (black dots) and H band intensities (grey
dots) plotted against the Ks-band surface brightness. The data have been
convolved to a resolution of 10 arc seconds and sampled at intervals of
FWHM/2.
}
\label{fig:plot_correlation_JHK}
\end{figure}

In this section we examine the relations between the three near-infrared
bands. Figure~\ref{fig:plot_correlation_JHK} shows the J- and H-band surface
brightness plotted against the Ks-band surface brightness. The data of
Fig.~\ref{fig:JHKmaps} was convolved to a resolution of 10 arc seconds using a
gaussian beam, while the plotted values correspond to a sampling of 5 arc
seconds. The zero point of the surface brightness values is set using the
reference area marked in Fig.~\ref{fig:JHKmaps}.

Figure~\ref{fig:plot_correlation_JHK} is qualitatively consistent with
the idea that the surface brightness is caused by light scattering. The
relations have a linear part and at higher Ks-band intensities (higher column
densities) saturation starts first at the shorter wavelengths. For the J-band
this happens below $I_{\rm Ks}=0.2$\,MJy/sr, whereas for the H-band the change
is more gradual and complete saturation is reached only after $I_{\rm
Ks}=0.3$\,MJy/sr. The expected optical depth in the J-band is about 1.7 times
the value in the H-band, so that the difference between the J- and H-bands is
of the expected magnitude. Although the dispersion increases towards higher
column densities, i.e. towards the centre of the filament, the NIR intensities
do still follow a relatively tight relation. Around $I_{\rm Ks}=0.5$\,MJy/sr
both J and H intensities show a hook, when the correlation with the Ks-band
becomes negative. Again, this is the expected behaviour for scattering in
optically thick clouds. Maximum surface brightness is reached when, for the
wavelength in question, the optical depth is around 1-1.5. Because H-band does
suffer complete saturation one can already estimate that, assuming a normal
extinction curve with $A_{\rm V}/A_{\rm H}\sim$6, the visual extinction must
exceed 10$^{\rm m}$ in a large part of the filament.

\begin{figure}
\resizebox{\hsize}{!}{\includegraphics{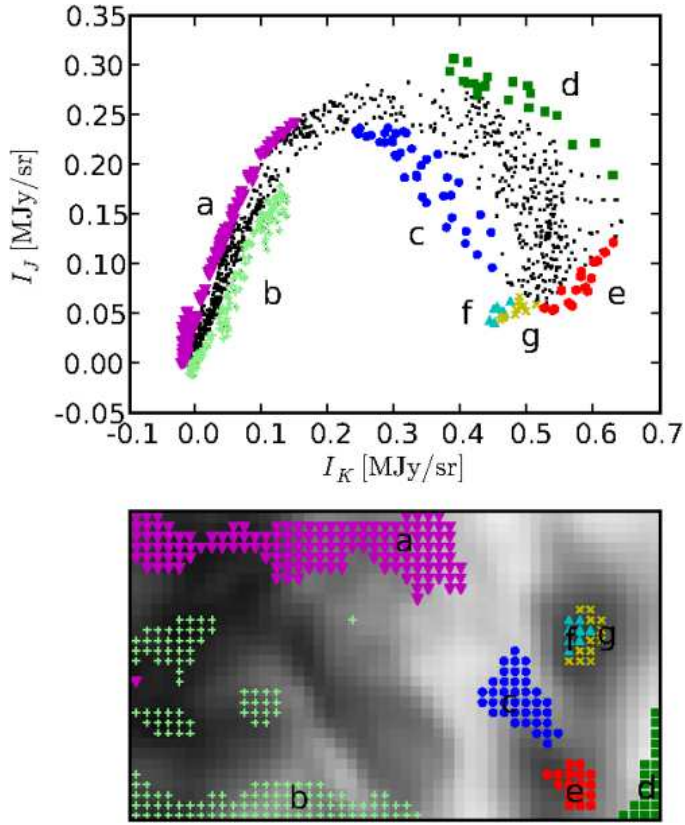}}
\caption{
{\em Upper frame:} J-band intensity plotted against the Ks-band intensity.
Some groups of measurements that deviate from the general trend are marked with
different symbols.
{\em Lower frame:} Map of the Ks-band intensity.  The groups of observations
identified in the upper frame are marked here with the same symbols. All data
have been smoothed to a resolution of 10$\arcsec$.
}
\label{fig:plot_correlation_map_JHK}
\end{figure}

If the surface brightness is to be used for column density estimation, there must
exist a single (although not necessarily linear) relationship between column
density and surface brightness. Therefore, we examine next the dispersion
around the mean relation. In Fig.~\ref{fig:plot_correlation_map_JHK} we show
again the relation between the J and Ks bands. This time we plot with
different symbols groups of measurements that deviate most from the mean
behaviour. In the second frame the distribution of these measurements is
overlaid on the J-band image. 

The linear part of the relation ($I_{\rm Ks}<0.15$MJy/sr) corresponds to
optically thin regions ($A_{\rm J}\la 1^{\rm m}$). Deviations from the average
relations could be caused by local changes in dust properties or variations in
the intensity or spectrum of the illuminating radiation.  However, when
optical depths are low the radiation sources should be very close before they
can produce significant intensity gradients.

Figure~\ref{fig:plot_correlation_map_JHK} shows that within the linear part
high J/Ks ratios are found to be systematically higher on the upper border
(region $a$) and lower on the lower border (region $b$) as compared to the
average. In between there are a few separate areas belonging to region $b$,
but these can be attributed to uncertainty in background subtraction (the
region that was used to define the zero level, see Fig.~\ref{fig:JHKmaps}) and
remaining artifacts around bright stars. The regions $a$ and $b$ can also be
identified in H/Ks images but are not visible in the J/H ratio  (see
Appendix \ref{sect:NIR_correlations}).  Therefore, the features must
originate in the Ks-band image. Moving upwards, the ratios J/Ks and H/Ks first
increase at the lower border, remain almost constant for a while, until in
region $a$ they increase again rapidly before a final drop at the upper
border. Even excluding borders affected by increased noise levels, the change
cannot be described as a smooth gradient across the image. One could
interpret, for example, region $a$ as a result of measurement and reduction
errors, e.g. imperfect flat fielding. However, the changes do partly coincide
with definite cloud structures. The dark area on the left of region $c$ and
above the rightmost tip of region $b$ is visible in J/H and H/Ks images as a
region of lower Ks-band emission. 

Above $I_{\rm Ks}\sim 0.2$\,MJy/sr saturation causes a turnover in the
relation between J and Ks intensities. The ratio of J and Ks optical depths
should be $\sim$2.5. Therefore, the relation between Ks intensity and column
density should still remain linear in these regions.
Figure~\ref{fig:plot_correlation_map_JHK} shows one distinct group of points
for which the J intensity is abnormally low. According to
Fig.~\ref{fig:plot_correlation_map_JHK}b this can be identified as a wedge of
low column density that runs into the filament (region $c$). One possible
explanation suggested by the geometry is that the shadowing provided by the
filament on both sides of the wedge has weakened the local radiation field,
the effect being strongest in the J-band.

The situation is reversed in the lower right hand corner of the field
(Fig.~\ref{fig:plot_correlation_map_JHK}, region $d$) which exhibits stronger
than average J band intensity. Based on the Ks band surface brightness the
column density should be similar as in region $c$. However, for a given
Ks-band intensity the value in the J-band can be higher by a factor up to
three. Note that for most of the cloud the points do follow a well-defined
relation which falls between the extremes represented by regions $c$ and $d$.
Region $d$ is on the other side of the optically thick filament where the
spectrum of the ISRF could be different. The average grain size could also be
smaller if, in that region, a shock wave has destroyed larger grains. On the
other hand, region $d$ is at the corner of the field and one cannot completely
rule out the possibility that it is merely an artifact.

As one moves to the left of region $d$, one enters the main filament and
region $e$, which coincides with a local minimum in the J intensity. Over this
short distance the Ks-band intensity remains roughly constant while in J the
intensity is reduced by a factor of $\sim$3--4. The minimum seen in
Fig.~\ref{fig:JHKmaps}a suggests that this is a local maximum of column
density and, while region $d$ is clearly outside the densest filament, region
$e$ represents the centre of it. However, according to Fig.~\ref{fig:JHKmaps}
in Ks-band the intensity peaks just SW of this position. This could mean
that within $e$ also the Ks band is saturated and, therefore, locally
anticorrelated with column density. This would be consistent with the
observation that the Ks-band intensity is still practically the same as in
region $d$, that is, in a region with seemingly much lower column density.
Assuming that the Ks-band has reached complete saturation, one can estimate the
optical depth to be above $\tau_{\rm Ks}\approx$1.5, or correspondingly the visual
extinction to be above $A_{\rm V}\approx$15$^{\rm m}$. Another explanation
would be that the Ks-band still follows the column density distribution, but
anisotropic illumination has shifted the J-band minimum towards the upper left. In
other words, the filament would be illuminated more strongly from the lower
right direction. 

The final two regions, $f$ and $g$, correspond to the most opaque region
because in Fig.~\ref{fig:JHKmaps}a the local minimum is seen at all three
wavelengths. Furthermore, in Fig.~\ref{fig:plot_correlation_map_JHK} these
form the end point in the non-linear relation between J and Ks intensities. By
dividing this tip into two parts we can see that even here points fall
systematically into separate locations. Region $g$ represents points that are
relatively brighter in Ks than in J. In the map
(Fig.~\ref{fig:plot_correlation_map_JHK}b) these form a crescent on the right
hand side (south) of the column density peak. Correspondingly on the left side
(north) the ratio J/Ks is larger. This suggests that the filament is
illuminated more strongly from the left. As one moves to the right, the intensity of the
illuminating radiation would become weaker and redder and, for a given column
density, the scattered radiation would not only be weaker but would also
exhibit a lower J/Ks ratio. The systematic shift in the location of the
intensity minimum was already visible in Fig.~\ref{fig:JHKmaps}.

The regions {\em c}--{\em g} can also be identified both from relations between H
and Ks bands and from the relation between J and H bands. This means that
it is unlikely that any of these features could have been produced by
measurement errors. On the other hand, the regions {\em a} and {\em b} are
probably largely artifacts. They may be indicative of the level of systematic
measurement errors that could be expected in observations of faint surface
brightness. These would ultimately set the limit to the accuracy at which
column densities can be determined based on surface brightness measurements.

\section{Column densities based on scattered light} \label{sect:conversion}

Padoan et al. (\cite{Padoan2006}) presented formulas for converting scattered
surface brightness into dust column density. At low visual extinctions the
relationship is linear and column densities could be mapped using observations
at any single wavelength. However, in order to obtain absolute values, some
assumptions must be made about the illuminating radiation field and the dust
properties. When the visual extinction reaches $A_{\rm V} \sim 10^{\rm m}$,
even NIR surface brightness values start to saturate. The effect begins at
shorter wavelengths and the relationship with column density becomes
increasingly more non-linear as the column density increases. In inhomogeneous
clouds with line-of-sight extinction $A_{\rm V}=10^{\rm m}$ the non-linearity
was found to be only $\sim$20\%. Padoan et al. (\cite{Padoan2006}) and Juvela et
al. (\cite{Juvela2006}) also discussed column density estimation using combined
measurements in J-, H-, and Ks-bands. In diffuse regions the intensity ratios
are determined by the spectrum of the external radiation field and by the dust
scattering properties. In more opaque areas the intensity ratios also depend
on the degree of saturation. In other words, the intensity ratios carry
independent information on the line-of-sight column densities. 

Following Padoan et al. (\cite{Padoan2006}) we first approximate the
relationship between surface brightness $I_{\lambda}$ and visual extinction
$A_{\rm V}$ with the equation
\begin{equation}
I_{\lambda} = a_{\lambda} \times ( 1 - e^{-b_{\lambda} A_{\rm V}} ).
\label{eq:a}
\end{equation}
Based on numerical simulations this functional form is approximately correct
as long as saturation remains weak. On the other hand, the equation cannot
describe strong saturation and the associated turnover in the relation between
intensity and column density. In the present case, this form remains
applicable outside the filament but it {\em cannot} be used to describe the J-
and H-band observations towards the centre of the filament. If J-, and H-bands
are compared with Ks-band data, $A_{\rm V}$ can be eliminated
\begin{equation}
I_{\lambda}  = 
a_{\lambda} \times 
( 1 - (1-I_{\rm K}/a_{\rm K} )^{b_{\lambda}/b_{\rm K}} ).
\label{eq:b}        
\end{equation}
The parameters $b_{\lambda}$ depend mainly on NIR dust properties and their
ratios can be assumed to be known. The parameters $a_{\lambda}$ can, in
principle, be obtained by fitting the Eq.~\ref{eq:a} to observations. Once
the parameters have been obtained, extinction estimates are derived from
the equation
\begin{equation}
A_{\rm V} = - \left[  log(1 - I_{\lambda}/a_{\lambda})  \right]  /   b_{\lambda}.
\label{eq:c}
\end{equation}
The final value of the column density can be calculated as a weighted sum of
estimates derived for J-, H-, and Ks-bands. 

In practice we fit to observations a parametric curve defined by
Eq.~\ref{eq:a}. The minimum distance between each observed triplet
($I_{J}$,$I_{H}$,$I_{Ks}$) and the curve is found by minimization. The sum
over squared distances gives an estimate of the goodness of fit, the
minimization of which gives values for the parameters $a$ and $b$. The
parameter $b_{\rm Ks}$ is kept constant, because, as seen in Eq.~\ref{eq:b},
observations can be used to fix only the ratio of the $b$-parameters. In the
fit the absolute value of $b_{\rm Ks}$ is unimportant, but it will determine
the final transformation into column density units. We use for $b_{\rm Ks}$ a
value of 0.125\,mag$^{-1}$ obtained from simulations made by Juvela et al.
(\cite{Juvela2006}). Those calculations were based on dust properties given in
Draine (\cite{Draine2003a}) and data files which are available on the web
\footnote{{\tt http://www.astro.princeton.edu/$~$draine/dust/}}. The assumed
ratio of total to selective extinction is $R_{\rm V}=A_{\rm V}/E(B-V)=3.1$.
The default extinction curve would predict $b_{\rm Ks}\sim$0.1\,mag$^{-1}$,
but, as discussed in Juvela et al. (\cite{Juvela2006}), $b$ cannot be
interpreted as a pure extinction cross section. 

\begin{figure} 
\resizebox{\hsize}{!}{\includegraphics{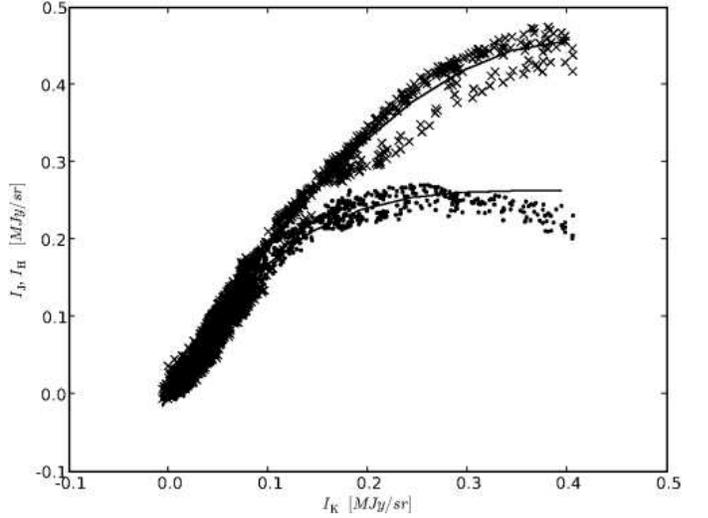}}
\caption{
The surface brightness in the J-band (dots) and H-band (crosses) plotted
against the Ks-band intensities. The lines show the fitted relations (see
text).
} 
\label{fig:JHKfit} 
\end{figure}

Figure~\ref{fig:JHKfit} shows the observed J- and H-band intensities values
plotted against the Ks-band intensities. The saturated part of the relations
($I_{\rm Ks}>0.4$\,MJy/sr) is not shown.  Here we avoid image borders, using
only data between declinations -36$\degr$52$^{\rm m}$30$^{\rm s}$ and
-36$\degr$50$^{\rm m}$0$^{\rm s}$. We also remove the remaining pixels
belonging to region $b$ (see Fig.~\ref{fig:plot_correlation_map_JHK}). The
resolution is still 10$\arcsec$. 
In the $H$-band one can see that not
all regions follow exactly the same relation, and there is a separate
population with low $H/Ks$ ratios. At high $I_{\rm Ks}$ these are close to
the already masked region $c$, and at lower $I_{\rm Ks}$ the points are associated
with emission close to the region $b$.  In Fig.~\ref{fig:JHKfit} we also show
the fitted analytic curve, projected into ($I_{\rm Ks}, I_{\rm J}$) and
($I_{\rm Ks}, I_{\rm H}$) planes. The fit is done using points with $I_{\rm
Ks}<0.4$\,MJy/sr. The limit corresponds to more than 20$^{\rm m}$ in visual
extinction.

For each observed point (J,H,Ks) the column density can now be estimated by
locating the nearest position on the curve, and using Eq.~\ref{eq:c}. The
selected value of $b_{\rm Ks}$ defines the final column density scale. The
effect of saturation decreases the correlation between intensity and column
density near the filament, especially in the $J$ band. On the other hand, at
low column densities the relative noise is largest in the Ks band.
Therefore, we adjust the relative weighting of the bands when estimating the
minimum distance from the curve. At low column densities the relative weights
are $J:H:Ks=1:0.8:0.2$, and they change linearly so that at the highest column
densities the weights are 0.2:0.8:1.6.

There is one additional complication that arises from the fact that the
reference area marked in Fig.~\ref{fig:JHKmaps}a is not free from extinction.
We could use Eq.~\ref{eq:a} and the surface brightness values relative to the
reference area and this way obtain estimates for extinction relative to the
reference area. However, a wrong zero point of the $A_{\rm V}$ axis could
cause a small bias because the relation between surface brightness and column
density is non-linear. The zero point is not important for Eq.~\ref{eq:a}
itself because that is only an analytic fit to observed data. However, we will
use a value of $b_{\rm K}$ that is based on simulations where zero intensity
does correspond to zero extinction. At low column densities, the intensity is
proportional to $a_{\rm K}\times b_{\rm K}$. An inaccurate value of $b_{\rm
K}$ is compensated by $a_{\rm K}$ that is obtained from a fit to observations.
The bias would exist mainly at high column densities. When one ignores
extinction in the reference region, the expected saturation of surface
brightness and the estimated column density will both be underestimated. Based
on 2MASS data and the NICER method (Lombdardi \& Alves \cite{Lombardi2001})
the minimum extinction in our field is $A_{\rm V} \sim 1.5^{\rm m}$. In the
next Section we use our NTT observations and the NICER method to obtain a more
accurate value, which is 1.8$^{\rm m}$ for the reference area. We use this
information already now, replacing Eq.~\ref{eq:a} with a corresponding
equation that takes into account the fact that zero intensity (relative to the
reference area) corresponds to an extinction $A_{\rm V}=A_{\rm ref}$,
\begin{equation}
I_{\lambda}(A_{\rm V})-I_{\lambda}(A_{\rm ref}) = 
a_{\lambda} \times ( e^{b_{\lambda} A_{\rm ref}} - e^{b_{\lambda} A_{\rm V}} ).
\label{eq:x}
\end{equation}
Another possibility would be to obtain the zero point for surface brightness
scale from OFF fields. However, the atmospheric emission and its variations
make it difficult to compare surface brightness levels between separate
fields. The magnitudes of the background stars are less affected by a changing
background level and provide a more reliable, albeit an indirect way to fix the
zero point of the extinction scale.

\begin{figure} 
\resizebox{\hsize}{!}{\includegraphics{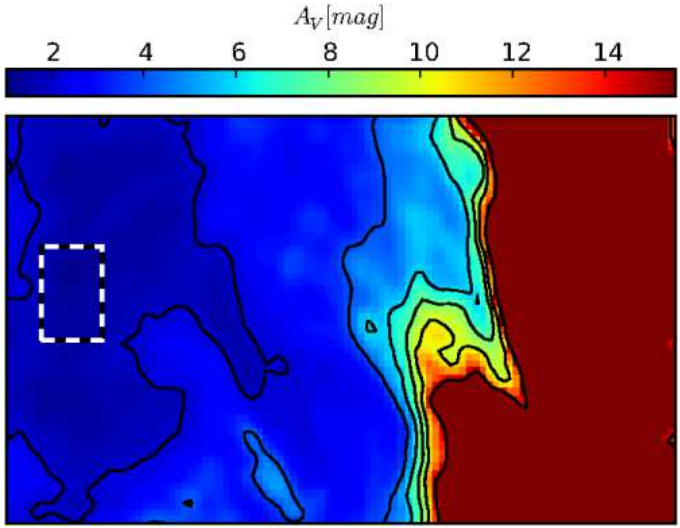}} 
\caption{
A map of column densities estimated based on the observed scattered surface
brightness and Eq.~\ref{eq:c}. The contours are drawn at $A_{\rm V}$ equal to
2, 4, 6, 8, 10, and 15 magnitudes. The resolution of the map is 10$\arcsec$
and the pixel size is 5$\arcsec$. For the reference area marked with the box
the extinction was assumed to be $1.8^{\rm m}$ (see text). No estimates are
shown above $A_{\rm V}=15^{\rm m}$ because Eq.~\ref{eq:c} becomes unreliable
at high extinctions.
} 
\label{fig:analytic_colden} 
\end{figure}

Figure.~\ref{fig:analytic_colden} shows a map of the estimated column
densities, here in units of visual extinction (as defined by the selected
value of $b_{\rm Ks}$). Column densities have been calculated for all pixels,
including, for example, also regions $a$ and $b$ (see above). The resolution
is 10$\arcsec$. When data below $I_{\rm Ks}=0.4$\,MJy/sr was fitted, we
obtained for $a_{\rm Ks}$ a value of 0.50\,MJy/sr. This is at the same time
the surface brightness that the analytical formulas predict for infinite
column density. The map does contain pixels where intensities are larger than
the value of the parameter $a$ and for which, therefore, no column density can
be estimated using this method. This is caused partly by observational noise
and partly because the analytic fit was done using only data from areas with
lower $A_{\rm V}$.  In the plot the $A_{\rm V}$ scale is cut at $15^{\rm m}$.
Above this limit the saturation becomes so strong that predictions of $A_{\rm
V}$ become unreliable. Equation \ref{eq:a} can clearly describe the surface
brightness only up to the point of turnover, and close to those values the
measurement errors are strongly amplified. At high $A_{\rm V}$ the results may
also be biased, because the formulas do not take into account shadowing caused
by the optically thick filament. Because of attenuation of the intensity of
the local radiation field the observed surface brightness decreases and, as a
result, the column density is underestimated. In more diffuse areas, e.g. in
regions $a$-$c$, the results can be affected by the deviating colours that
were discussed above.

\section{Extinction map from background stars} \label{sect:NICER}

We use observations of the background stars and the NICER algorithm (Lombdardi
\& Alves \cite{Lombardi2001}) to derive an extinction map for the observed
field. NIR images give colours for a large number of stars that reside behind 
the Corona Australis cloud and whose radiation suffers reddening because of
the dust in the cloud. Visual extinction can be estimated by comparing these
reddened colours with the colours observed towards an unextincted off--region.
The method relies on the assumption that, in a statistical sense, the
intrinsic colours of the stars are similar in the off--region and in the
region studied. Knowledge of the dust extinction curve is needed in order to
combine the information of several bands and in order to transform colour
excesses into values of visual extinction, $A_{\rm V}$.

\begin{figure} 
\resizebox{\hsize}{!}{\includegraphics{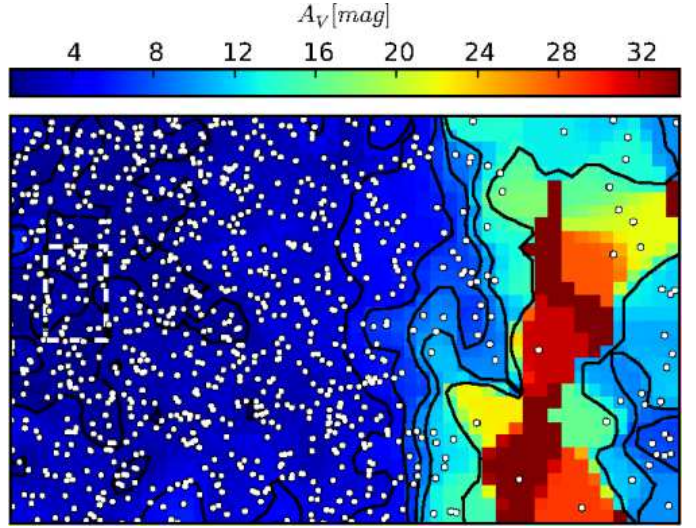}} 
\caption{
Extinction map estimated based on the reddening of background stars. The
contours are drawn at levels of 2, 4, 6, 8, 10, and 15 magnitudes of $A_{\rm
V}$. The resolution is $20\arcsec$ and the pixel size is 10$\arcsec$. The
stars used in the calculation are marked as dots. The average extinction
within the marked reference area is $A_{\rm V}=1.8^{\rm m}$.
} 
\label{fig:extinction_map} 
\end{figure}

In the following we assume a normal extinction curve (Cardelli et al.
\cite{Cardelli89}). Analysis of 2MASS data that employed a dark off-region
well outside the cloud suggested that there is significant extinction
everywhere in our ON fields. Therefore, in order to obtain an absolute zero
point for the $A_{\rm V}$ scale we use the field OFF-1 as the reference
region. In the ON fields, even with our deep NIR observations, the stellar
density is too low to be able to determine the extinction at a resolution of
$10\arcsec$. The number of stars for which magnitudes could be measured in all
three bands is less than 1000 for the whole map, and the stellar density
decreases rapidly as one moves into the optically thick filament. Therefore,
in Fig.~\ref{fig:extinction_map} we show an extinction map calculated at a
resolution of 20$\arcsec$. 
The sigma clipping was done at a 3-$\sigma$ level (see Lombdardi \& Alves
\cite{Lombardi2001}).
The stars used for the extinction measurement are plotted in the same figure.
Note that although the extinction map does contain a region with predicted
extinction $A_{\rm V}>20$, that region contains very few stars. 
For the reference area (see Fig.~\ref{fig:JHKmaps}a) used in
surface brightness plots we obtain an average extinction of $A_{\rm
V}=1.8^{\rm m}$. This takes into account the estimated extinction within the
OFF fields (see Table~\ref{table:fields}). This value was already used in the
previous Section, in the conversion between surface brightness and $A_{\rm
V}$.

\section{Comparison of column density estimates} \label{sect:comparison}

In this section we compare in more detail the column density estimates
obtained from scattered light with those derived from background stars.
We limit the comparison to areas where the scattered light predicts values
$A_{\rm V}<15^{\rm m}$. The limit is set because the methods of
Sect.~\ref{sect:conversion} are no longer applicable when Ks band surface
brightness suffers significant saturation. 
%% MJ
In this Section we examine only column density estimates north of the filament
(left of the filament in the orientation used, for example, in
Fig.~\ref{fig:JHKmaps}). In Sect.~\ref{sect:conversion} the parameters were
determined using only that area. On the other side of the optically thick
filament the radiation field could be different, and we return to this
question later in Sect.~\ref{sect:discussion}.

\subsection{Comparison of extinctions} \label{sect:ext_comparison}

Some conclusions can be drawn already by looking at
Figs.~\ref{fig:analytic_colden} and \ref{fig:extinction_map}. The maps agree
in their main features and give even quantitatively similar $A_{\rm V}$
estimates. In spite of the fact that Fig.~\ref{fig:analytic_colden} has twice
the resolution of the NICER map it is significantly smoother. This strongly
suggests that even at the resolution of $20\arcsec$ the NICER map still
contains significant noise caused by the scatter in the intrinsic colours of
the background stars. In many places where the details of the two $A_{\rm V}$
maps differ, the differences can be attributed to low stellar density.  For
example, in the lower central part of Fig.~\ref{fig:analytic_colden}, the
surface brightness data reveal an elongated region with dimensions of $\sim
90\arcsec \times 30 \arcsec$ that in this figure is surrounded by a contour
at $A_{\rm V}=4^{\rm m}$. The feature should be resolved at 20$\arcsec$
resolution, but the NICER map contains only a hint of this feature because
there happens to be no background stars within this area.  The wedge of low
extinction that extends into the filament is much narrower in the map that is
based on surface brightness data. This is not caused by the difference in the
selected resolution but rather by the fact that also this area contains only a
few background stars. The surface brightness data also predict a sharp
increase of extinction close to the filament. This is probably real although a
gradient could be altered if the parameters used in the analysis were wrong.
For example, if parameter $b_{\rm K}$ were overestimated, one would expect
high surface brightness values to be strongly saturated and the higher the
intensity the more would the column density be overestimated. The presence of
such errors cannot be confirmed or refuted by the NICER data. There the
15$^{\rm m}$ contour is certainly further away from the lower contours but
stellar density is again too low for this to be significant. The NICER map was
created using all stars where we had photometric values for all three bands.
The outcome is not improved when the map is made using all stars detected in
two or three bands. The number of stars is naturally increased but at the same
time the errors for individual stars increase.

\begin{figure} 
\resizebox{\hsize}{!}{\includegraphics{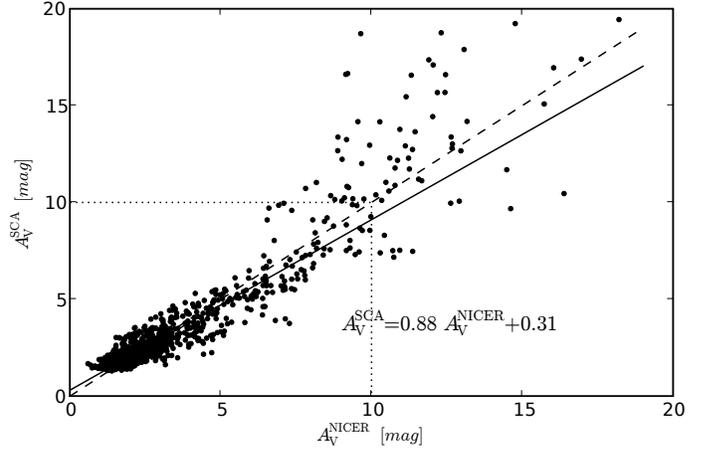}} 
\caption{
Correlation between the extinction derived using the NICER method, $A_{\rm
V}^{\rm NICER}$, and the values derived from surface brightness measurements,
$A_{\rm V}^{\rm SCA}$. The comparison is limited to area left of the
filament (see Fig.~\ref{fig:analytic_colden}), where the surface brightness
method predicted extinctions below 20$^{\rm m}$. The points correspond to
pixels in Fig.~\ref{fig:extinction_map}: the resolution is 20$\arcsec$ and the
sampling is done at intervals of 10$^{\arcsec}$.
} 
\label{fig:ext_correlation} 
\end{figure}

In the following the comparison is carried out at the resolution of
20$\arcsec$, because of the lower resolution of the colour excess maps. 
Figure~\ref{fig:ext_correlation} shows the resulting correlation between the
column density estimates based on scattered light, $A_{\rm V}^{\rm sca}$, and
those derived using the NICER method, $A_{\rm V}^{NICER}$. 
The plot is limited to the region with $A_{\rm V}<20^{\rm m}$ and,
therefore, includes the regions $a$--$c$ discussed in
Sect.~\ref{sect:qualitative}. The pixels belonging to the areas $a$ and $b$ do
not deviate from the general relation. In the range $A_{\rm
V}^{NICER}=8-11^{\rm m}$ most of the points above the average relation
correspond to the region $c$.
In the plot the two $A_{\rm V}$ scales are related, but only indirectly. For
NICER the absolute scale is defined by the assumed extinction law (Cardelli et
al. \cite{Cardelli89}). For scattered light the scaling depends on the
selected value of $b_{\rm Ks}$. This was obtained from simulations that
assumed a dust model consistent with a similar extinction law (Draine
\cite{Draine2003a}). However, while the NICER estimates depend only on total
dust extinction cross sections, the observed intensity of scattered light also
depends on the scattering function and albedo of dust grains. In
Fig.~\ref{fig:ext_correlation} the extinction estimates are consistent with
each other, and, in first approximation, $A_{\rm V}^{\rm sca} \sim A_{\rm
V}^{NICER}$ (dashed line in Fig.~\ref{fig:ext_correlation}). This can be taken
as further confirmation that most of the observed surface brightness is indeed
caused by scattering from dust particles, because the values of $A_{\rm
V}^{\rm sca}$ do explicitly depend on this assumption and the ratio between
extinctions in the $Ks$ and $V$ bands. The ratios between constants $b$ and
the values of the constants $a$ at J, H, and Ks were obtained directly from
the data. Therefore, we did not need to assume any particular intensity or
spectral shape for the illuminating radiation. The only assumption is that
these remain constant over the mapped area.

At low $A_{\rm V}$ the scatter increases because of measurement errors. The
extinction maps of Figs.~\ref{fig:analytic_colden} and
\ref{fig:extinction_map} already indicated that the scatter would be dominated
by uncertainty in the NICER method.  In the colour excess method the
uncertainty has two components, one arising from photometric measurement
errors and another from the scatter in the intrinsic colours of background
stars. 
For a more thorough discussion of the statistical basics of the colour
excess method see Lombardi (\cite{Lombardi2005}). That paper also describes the
combination of the colour excess and star count methods first discussed by
Cambr\'esy et al. (\cite{cambresy2002}).
In our case the formal errors are dominated by variation in the
intrinsic colours. However, the error estimates reported by the NICER program
do not take into account true opacity variations within the beam that is used
for spatial averaging of the extinction values of individual stars. If it is
estimated based on this dispersion, the error increases with increasing
extinction. At high $A_{\rm V}$ there can also be a significant bias because
background stars are observed preferentially from those parts of the cell
where extinction is lower. In Fig.~\ref{fig:ext_correlation}, at high
extinctions, the surface brightness data tend to predict larger $A_{\rm V}$
values than what the background stars do. This can be a direct result of the
low number of stars seen through the filament. In the NICER maps and
especially in the region $A_{\rm V}\sim 10-15^{\rm m}$ many values are
obtained by interpolating over very long spatial distances. In the case of the
filament, this can result in significantly underestimated extinction values.
The scatter in the $A_{\rm V}^{\rm NICER}$ values carries information on
column density variations within the cells that are used in the calculation of
the extinction map (Padoan et al. \cite{Padoan1997}; Juvela \cite{Juvela1998};
Lada et al. \cite{Lada1999}). However, in our case those variations are
dominated by the large scale gradients associated with the filament rather
than any small scale inhomogeneities.

\subsection{Comparison of errors}

\begin{figure} 
\resizebox{\hsize}{!}{\includegraphics{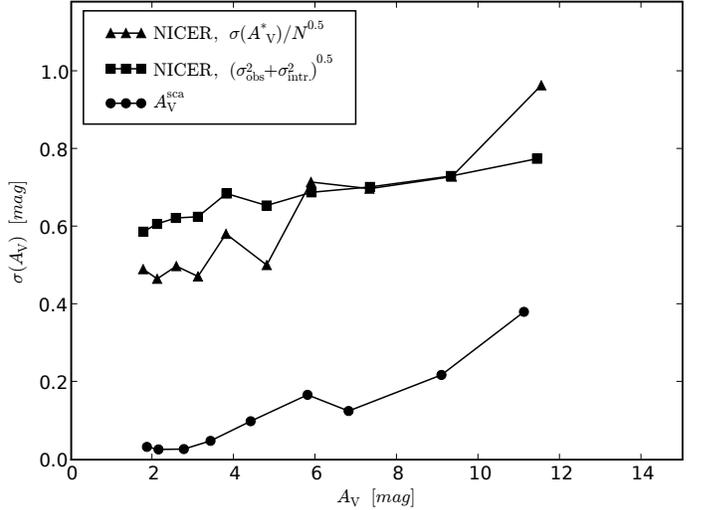}} 
\caption{
Estimated uncertainty of extinction estimates as a function of $A_{\rm V}$.
For the surface brightness method the dispersion in estimates of $A_{\rm V}$
is shown as solid circles. For NICER the dispersion of intrinsic colours was
estimated from the field OFF--1. Adding photometric errors reported by
DAOPHOT, the solid squares represent formal error of NICER predictions. The 
triangles show the error of the mean for stars within each
20$\times$20$\arcsec$ grid cell (minimum of 2 stars per cell). Thus these
values also reflect the $A_{\rm V}$ variations within the cells (gradients
{\em and} small scale structure) and the sampling provided by background
stars. 
} 
\label{fig:errors} 
\end{figure}

In Fig.~\ref{fig:errors} we plot two error estimates for the NICER method
calculated for $20\times20\arcsec$ cells. The first error estimate is obtained
by combining the effects of photometric errors, $\sigma_{\rm obs}$, and the
scatter in the intrinsic colours of the stars, $\sigma_{\rm intr}$. The error
in $A_{\rm V}$ values was estimated by determining the variation in extinction
maps, when noise was added to the input data. For the errors of input
magnitudes we use the DAOPHOT error estimates. The scatter of intrinsic
colours was estimated from an OFF field and this dominates the overall
uncertainty. For individual stars the estimated error is $\sigma(A_{\rm
V})\sim 1.5^{\rm m}$ almost irrespective of the extinction. At low $A_{\rm V}$
the 20$\times 20\arcsec$ cells contain typically $\sim$10 stars.

The second error estimate is obtained by calculating the scatter
of the observed $A_{\rm V}$ values of individual stars in a cell and by
dividing this number with the square root of the number of stars.  For a flat
$A_{\rm V}$ map the two error estimates should be equal provided that the
errors $\sigma_{\rm obs}$ and $\sigma_{\rm intr}$ are correctly determined.
However, the second error estimate also includes some information on the
column density variations within the map cells. 
If there are strong extinction gradients it could overestimate the true error.
On the other hand, it does not take into account the bias that exists because
of the anticorrelation between the column density and the surface density of
the observed background stars.
Figure.~\ref{fig:errors} shows that the error estimates obtained with the
second method increase more rapidly as extinction increases. This is probably
caused by the steep extinction gradients. The plot extends only up to $A_{\rm
V} \sim 12^{\rm m}$ because, at higher extinctions, there are very few cells
with more than one star per cell.

In the case of scattered light the estimation of the uncertainty of the
surface brightness measurements is not quite straightforward because the
images also contain true surface brightness variations at all scales and we do
not know how the errors are correlated over different distances. In order to
remove real large scale structures, we filtered out all Fourier terms
corresponding to spatial frequencies $k<0.1$\,arcsec$^{-1}$. One could
estimate the uncertainty of surface brightness by calculating the remaining
standard deviation between individual pixels. However, if the noise is
correlated over several pixels, this will underestimate the true errors.
Instead, we divided each 20$\arcsec \times 20 \arcsec$ cell into $4\times 4$
sub-regions and used their dispersion to estimate the error of the mean. The
results (circles in Fig.~\ref{fig:errors}) should be independent of correlated
errors at scales below 5\,$\arcsec$.
%% MJ
The estimates still include the effect of true column density variations at
this scale and, therefore, could overestimate the true uncertainty. 

The Fig.~\ref{fig:errors} indicates that, in our case, the uncertainty in
$A_{\rm V}$ is smaller for the surface brightness method than for the colour
excess method. The previous analysis does not fully take into account the bias
of the estimates. Nevertheless, the results suggest that the sampling errors
caused by the limited number of background stars have a significant
contribution to the scatter seen in Fig.~\ref{fig:ext_correlation}, especially
at high extinctions.
Lombardi (\cite{Lombardi2005}) described a maximum likelihood method that,
in the presence of foreground stars, showed smaller bias than the NICER
method. The use of such schemes might improve the results of the colour excess
method also here but, of course, cannot overcome the basic problem of an
insufficient surface density of the background stars.

\section{Discussion} \label{sect:discussion}

Our main goal was to confirm that observations of scattered surface brightness
can be used to map the column density of quiescent clouds. The comparison with
extinction estimates calculated with the NICER method shown in
Fig.~\ref{fig:ext_correlation} revealed a good correlation, especially below
$A_{\rm V}\sim 10^{\rm m}$.

\subsection{Contamination by dust emission}

\begin{figure} 
\resizebox{\hsize}{!}{\includegraphics{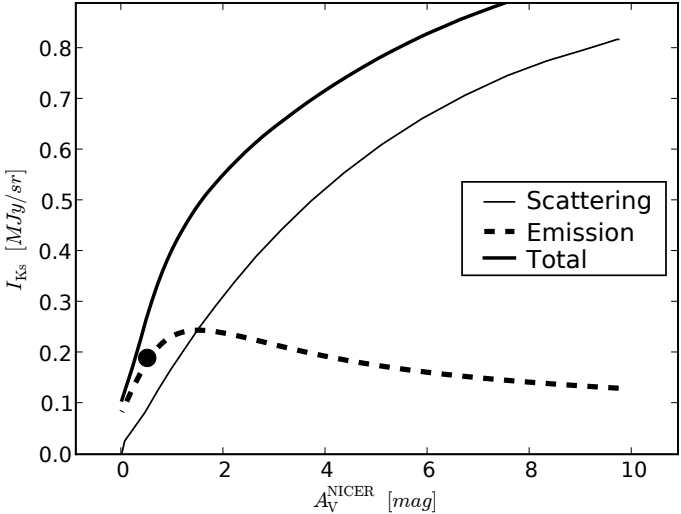}} 
\caption{
Predicted distribution of scattered and emitted radiation in the case of
a spherically symmetric cloud with $A_{\rm V}=10^{\rm m}$ for the sightline
through the cloud centre. The model is based on the Mathis et al.
(\cite{Mathis83}) radiation field and the dust model described in
Sect.~\ref{sect:conversion}. The emission has been scaled to correspond to the
Flagey et al. (\cite{Flagey2006}) prediction calculated for $A_{\rm
V}=0.5^{\rm m}$.
} 
\label{fig:emission} 
\end{figure}

At low column densities a significant fraction of the surface brightness
could, in principle, be caused by dust {\em emission} (for discussion, see
Juvela et al. \cite{Juvela2006}). This applies mainly to observations in the
Ks band whereas in the H and J bands dust emission should be insignificant
compared with light scattering. The good agreement between column densities
estimated using background stars and using the surface brightness suggests
that the level of emitted signal is low enough not to interfere with the
estimation of column densities. The same conclusion could, in principle, be
reached more directly by comparing surface brightness values or by plotting
the Ks surface brightness against $A_{\rm V}$ obtained independently from
background stars. Possible emission is limited to Ks band and to outer cloud
layers and, therefore, emission would cause non-linearity in these relations.
The relation between surface brightness values was shown in
Fig.~\ref{fig:JHKfit}. There is no sign that the slopes of J and H vs. Ks
would get steeper towards lower $A_{\rm V}$. In Fig.~\ref{fig:emission} we
show the predicted spatial distribution of scattered signal and dust emission
for a spherical model cloud where extinction through the centre of the cloud
is $A_{\rm V}=10^{\rm m}$. The model uses the dust model mentioned in
Sect.~\ref{sect:conversion}. However, in this plot the intensity of emission
has been scaled up to correspond to results of Flagey et al.
(\cite{Flagey2006}), which predict at 2$\mu$m an emission of $\sim$0.03\,MJy\,
sr$^{-1}$ corresponding to $N_{\rm H}=10^{21}$\,cm$^{-2}$ (see their Fig. 9).
In Fig.~\ref{fig:emission} the emission curve is scaled so that it goes
through the corresponding point calculated for $A_{\rm V}=0.5^{\rm m}$. Note
that according to the original Li \& Draine (\cite{Li2001}) dust model the
emission in K band would be only 20\% of the scattered signal. In
Fig.~\ref{fig:JHKfit} the intensities were given in relation to a reference
region where extinction was $A_{\rm V}=1.8^{\rm m}$. According to
Fig.~\ref{fig:emission} this is the regime where emission is expected to be
the strongest. Therefore, by taking the difference with respect to the
reference, most of the effect of possible emission has already been eliminated
from this plot. The existence of emission could be tested sensitively only if
the NIR observations extended well below $A_{\rm V}=1^{\rm m}$.

\begin{figure} 
\resizebox{\hsize}{!}{\includegraphics{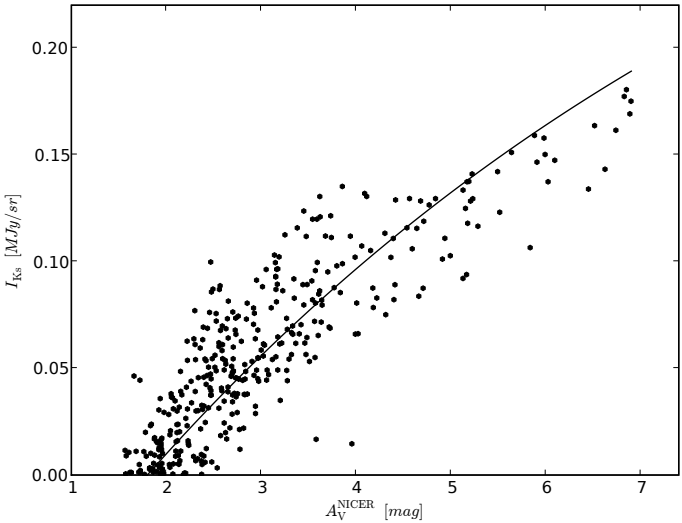}} 
\caption{
The observed $Ks$ intensity plotted against the extinction obtained with the
NICER method. The intensities are relative to the value found in the reference
region (see Fig.~\ref{fig:JHKmaps}). Only points below $A_{\rm V}=7^{\rm m}$ are
shown. The curve indicates the expected curvature based on numerical modelling.
} 
\label{fig:correlation_AV_scat}
\end{figure}

In Fig.~\ref{fig:correlation_AV_scat} we plot observed Ks intensities against
extinctions calculated with the NICER method. The plotted curve represents the
expected curvature of this relation (i.e., it corresponds to the value of
$b_{\rm K}$ quoted above). The value was obtained from numerical modelling
assuming normal dust properties. In the case of dust emission clouds should
show some degree of limb brightening that, in this figure, would cause the
slope to become shallower at low extinctions. There is no indication of such a
trend, neither in the internal distribution of data points nor in comparison
with the predicted behaviour due to scattering only. If the emission were as
strong as depicted in Fig.~\ref{fig:emission}, the signal at $A_{\rm
V}=1.5^{\rm m}$ would be increased by almost 0.1\,MJy/sr compared with the
total signal at $A_{\rm V}=6^{\rm m}$. Based on
Fig.~\ref{fig:correlation_AV_scat} the contribution of dust emission must be
lower by almost a factor ten. Note, however, that the prediction of
Fig.~\ref{fig:emission} is based on a simple, spherically symmetric cloud with
a smooth density distribution. For example, inhomogeneities or a different
cloud geometry could well decrease the contrast between cloud edge and
sightlines of higher extinction. Nevertheless, it is safe to say that dust
emission is unlikely to interfere with the use of surface brightness as a
tracer of cloud column densities.

\subsection{Sampling effects in the NICER extinction maps}

\begin{figure*} 
\resizebox{\hsize}{!}{\includegraphics{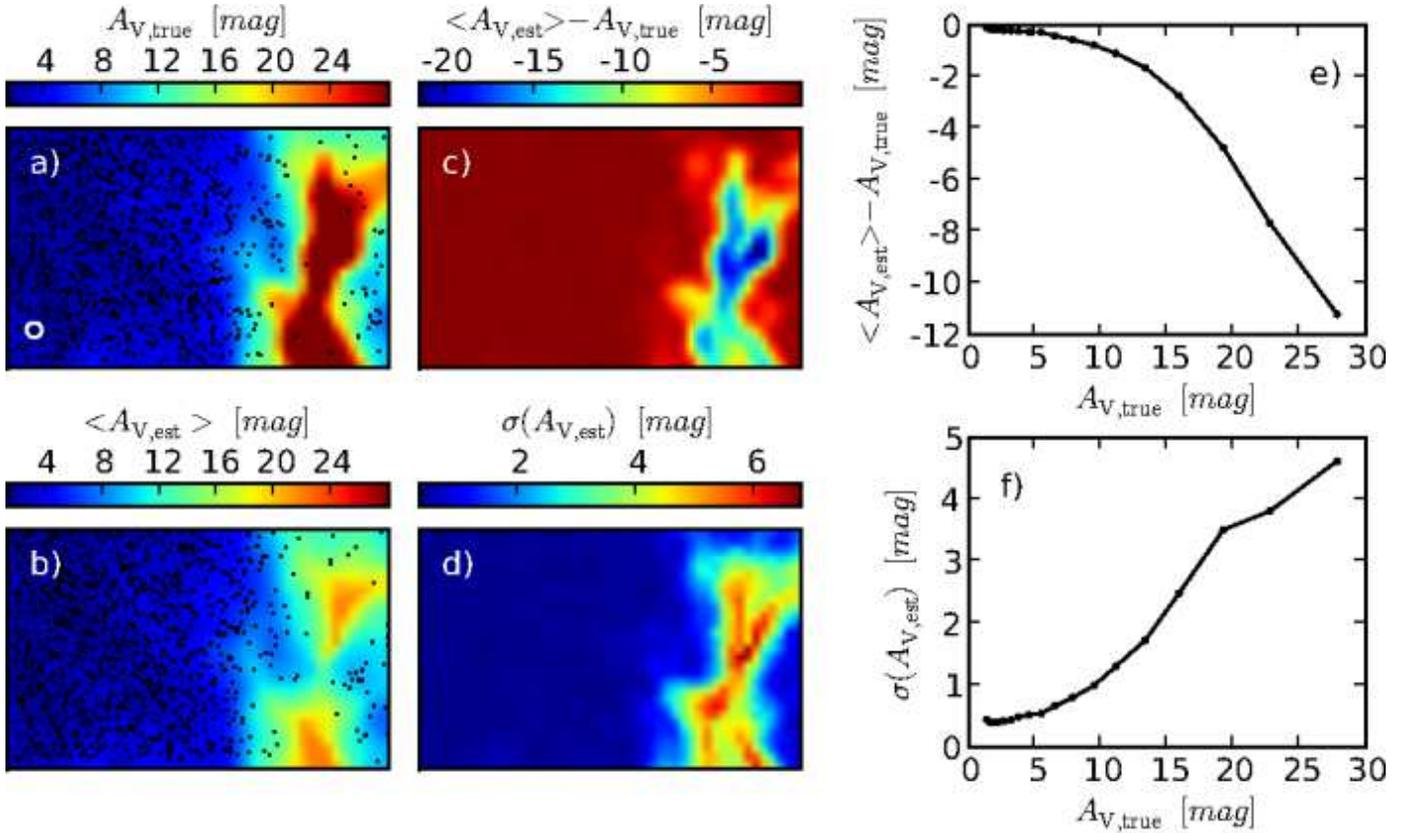}} 
\caption{
Simulations of estimated errors in the NICER $A_{\rm V}$ map.  The $A_{\rm V}$
distribution obtained with the NICER method (frame $a$) is used as a starting
point. New $A_{\rm V}$ maps are obtained by simulating stars at random
locations, frame {\em b} showing the average of these. The observed stars (frame
{\em a}) and an example of simulated distribution of stars (frame {\em b}) are
indicated with black dots. The other images show the average bias (frame $c$), and
internal dispersion (frame $d$) of the $A_{\rm V}$ estimates. The rightmost
plots show the bias and the internal standard deviation of the $A_{\rm V}$
estimates as functions of true visual extinction. All maps correspond to
averages weighted with a gaussian with FWHM equal to 20$\arcsec$ (the circle
in frame {\em a}).
} 
\label{fig:sampling_errors} 
\end{figure*}

We noted already that the NICER values may have been affected by poor sampling
of regions with strong extinction gradients. We examined this possible bias
with simulated observations. As a starting point we take the map $A_{\rm
V}^{\rm NICER}$ which is first interpolated to a ten times higher resolution.
Background stars are simulated using the magnitude and intrinsic colour
distributions obtained from observations. The simulations take into account
the observed, two-dimensional probability distribution of the colours. The
synthetic observations are run through the NICER routine resulting in a new
extinction map. The procedure is repeated so that we obtain 100 realizations
of the extinction map. In Fig.~\ref{fig:sampling_errors} the results are
compared with each other and with the input $A_{\rm V}$ distribution, again
using the 20$\arcsec$ resolution.

Frame {\em a} shows the original $A_{\rm V}$ distribution and the positions of
the observed stars. The average recovered map (frame {\em b}) follows the true
extinction distribution but within the filament the recovered extinction
values are systematically too low. This is not surprising as the number of
stars observed in this region is very small. The bias is plotted in frame {\em
c}. The largest errors are associated with opacity gradients in optically
thick regions. At the centre of the filament the errors can again decrease, as
the $A_{\rm V}$ distribution flattens out. In Fig.~\ref{fig:sampling_errors}c
this is true especially at the lower part of the map. Frame {\em d} shows
the dispersion between different realizations of the extinction map. The map
shows a similar pattern, statistical errors increasing where also the
systematic errors are large. The last two plots in
Fig.~\ref{fig:sampling_errors} show the bias and scatter of the $A_{\rm V}$
estimates as functions of true extinction. Both systematic and statistical
errors increase monotonically with $A_{\rm V}$, although it is probably
significant that in our case large $A_{\rm V}$ also often corresponds to 
large $A_{\rm V}$ gradients. The plots extend to regions with true extinction
above 25$^{\rm m}$. However, as seen from frames {\em a} and {\em b}, above
$A_{\rm V} \sim 15^{\rm m}$ the estimates mostly correspond to interpolated
values between stars more than 20$\arcsec$ apart.

Figure~\ref{fig:sampling_errors} indicated a significant bias with respect to
the input map. Therefore, one can assume that also the original NICER
estimates contain similar errors. For example, when the estimated $A_{\rm V}$
value is around 15$^{\rm m}$, the true value could well be above $20^{\rm m}$.
In fact, the errors could be even slightly larger, because our simulation does
not include the effect of small scale opacity variations. Therefore, this bias
could explain most of the discrepancy that was seen in
Fig.~\ref{fig:ext_correlation}. When extinction derived from surface
brightness data is around $15^{\rm m}$, the NICER method predicts values that
are lower by $\sim 5^{\rm m}$.

\subsection{Interstellar radiation field and variations of dust illumination}

Remarkably, the surface brightness data could be converted into column
densities without any assumption on the intensity or spectral shape of the
local radiation field. This was because most parameters, $a_{\rm J}$, $a_{\rm
H}$, and $a_{\rm Ks}$ in particular, were obtained by fitting Eq.~\ref{eq:a}
directly to observations. Only the parameter $b_{\rm Ks}$ was fixed to the
value obtained from earlier model calculations. The parameter $b_{\rm Ks}$
represents mainly the Ks band optical depth and, therefore, should be independent of
the intensity of the radiation field. In Juvela et al. (~\cite{Juvela2006})
$a_{\rm Ks}$ and $b_{\rm Ks}$ were fitted simultaneously and, because the
functional form is not perfect in describing the observed relation between
surface brightness and column density, a change in the radiation field could
slightly affect also the parameter $b_{\rm Ks}$. The effect should be small
and, therefore, we can assume that the column density estimates presented in
this paper are effectively independent from the radiation field assumed in
Juvela et al.(~\cite{Juvela2006}).

\begin{figure} 
\resizebox{\hsize}{!}{\includegraphics{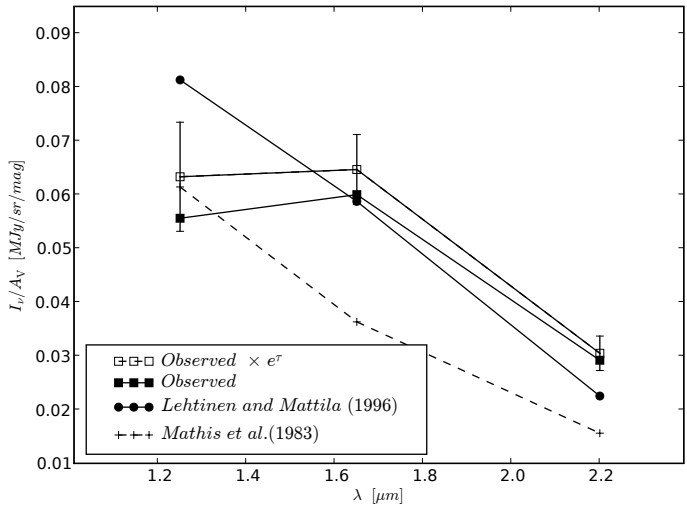}} 
\caption{
Observed and predicted intensities of scattered light per unit visual
extinction. The observed spectrum (filled squares) corresponds to fits to
surface brightness against NICER estimated visual extinction. The uppermost
curve (open squares) is obtained by multiplying these with
$e^{\tau}$, where $\tau$ corresponds to the expected attenuation surrounding
the filament.
The other curves show spectra of scattered light assuming either Mathis et al.
(\cite{Mathis83}) or Lehtinen et al. (\cite{Lehtinen1996}) radiation field, and
normal dust scattering properties (see text).
} 
\label{fig:check_ISRF} 
\end{figure}

If the ratios between the scattering cross sections of all bands are assumed
to be known, observations of optically thin sightlines give the spectrum of
the illuminating radiation field. Because we have already estimated visual
extinctions, we can estimate also the absolute intensities. In
Figure~\ref{fig:check_ISRF} we compare the observed surface brightness values
with predictions, assuming the normal dust properties used in this paper (see
Sect.~\ref{sect:conversion}). For the observations we plotted surface
brightness against NICER estimates of $A_{\rm V}$. Linear fits were made to
data where optical depth in the corresponding band was below 0.7.
Figure~\ref{fig:check_ISRF} shows the slopes, which represent the intensity of
the illuminating radiation field. We should further take into account the fact
that in our ON field the minimum $A_{\rm V}$ value is $\sim$1.8 magnitudes. If
the cloud is surrounded by an isotropic attenuating layer, the true ISRF
outside this layer should be higher by a factor of $e^{\tau_{\nu}}$, where
$\tau_{\nu}$ is the corresponding NIR optical depth. In reality the effect is
decreased because of the strong forward scattering in the NIR. We assume
effective $\tau_{\nu}$ values corresponding to $\sim$0.5 magnitudes of visual
extinction. The resulting ISRF estimates are shown in
Fig.~\ref{fig:check_ISRF} with open squares. The error bars have been
calculated with the bootstrap method.

In Fig.~\ref{fig:check_ISRF} we plot also slopes corresponding to the Mathis
et al. (\cite{Mathis83}) radiation field, i.e., the values obtained by
multiplying the given Mathis et al. intensity estimates with the scattering
cross sections of the Draine dust model. Based on COBE/DIRBE NIR data Lehtinen
\& Mattila (\cite{Lehtinen1996}) presented improved estimates of NIR ISRF.
Compared with Mathis et al. those ISRF values and, consequently, the expected
surface brightnesses are larger by roughly one half.

The spectrum of scattered light in Corona Australis is rather similar to what
is expected for a cloud illuminated by normal ISRF. The H and Ks intensities
are larger than the values given by Lehtinen \& Mattila. However, the
difference is significant only in the Ks band where the observed value is
about 35\% ($\sim$ 2\,$\sigma$) larger. This cannot be an indication of
additional component of dust emission in that band, because the spectrum is
based on differential measurements of surface brightness in regions with
$A_{\rm V}\sim 1.8^{\rm m}$ or higher. Dust emission would peak around the
lowest $A_{\rm V}$ values and, therefore, would decrease rather than increase
the slope in the relation of surface brightness vs. visual extinction. 
Between the J and Ks bands the shape of the observed spectrum is very close to
both ISRF models. The J band intensity is relatively lower and quite close to
the Mathis et al. value. Because of the larger error estimates, the difference
to the Lehtinen \& Mattila value amounts only to 1.5\,$\sigma$.  A correction
$e^{\tau_{\nu}}$ corresponding to a $A_{\rm V}=1.5^{\rm m}$ would be enough to
remove this difference.  A larger correction would be justifiable, because the
filament is optically thick and blocks practically all incoming radiation from
that direction. Therefore, the shape of the ISRF may be close to that
predicted by the models mentioned. However, with the possible exception of the
J band the intensity is higher than predicted by the two ISRF models. Had we
used in the correlations $A_{\rm V}$ values estimated based on the surface
brightness instead of the NICER estimates, the intensity difference would
further increase by $\sim$10\% (see Fig.~\ref{fig:ext_correlation}). 

In Fig.~\ref{fig:check_ISRF} the ISRF models were converted into expected
surface brightness of scattered light using the Draine dust model. In the
model the albedos are close to the lower limit of the range that Lehtinen \&
Mattila (\cite{Lehtinen1996}) derived from NIR observations. In
Fig.~\ref{fig:check_ISRF} larger albedo values would move the Mathis et al.
and the Lehtinen \& Mattila ISRF curves upwards in the same proportion. In the
NIR the scattering is also mostly in the forward direction and the intensity
of the scattered light should depend more on the radiation coming in from
behind the cloud than on the average intensity over the whole sky. In the
Draine dust model the asymmetry parameter of scattering, $<cos \theta>$, is
0.28 for the J band and 0.13 for the K band. Using the DIRBE zodiacal light
subtracted all-sky maps at 1.25\,$\mu$m and 2.2\,$\mu$m one can estimate that
the scattered surface brightness of the Corona Australis cloud should be
20-30\% higher than what would be expected based on the average sky brightness
alone. Therefore, the strong forward scattering can also explain
why the ISRF estimate in Fig.~\ref{fig:check_ISRF} is above the two models.

The star R Corona Australis lies only some five arc minutes East of our field.
In the visual (see Fig.~\ref{fig:plot_dss}) the associated reflection nebula
does not extend near our ON field, but in NIR the star could have a
significant effect at that distance. The intervening extinction cannot be
determined. If the star is sufficiently far in the foreground, it could
illuminate also the Northern side of the filament. In the 2MASS catalog R
Corona Australis has observed magnitudes of 6.9, 5.0, and 2.9 in J, H, and K,
respectively (Cutri et al. \cite{Cutri2003}). Assuming a distance of 130\,pc
(Marraco \& Rydgren \cite{Marraco81}) the projected distance between R CrA and
the centre of our field is $\sim$0.5\,pc. If we assume that the extinction
between R CrA and the observed field is similar to the extinction between
between R CrA and us, the resulting ratio of energy densities produced by
R\,CrA and the Mathis ISRF would be 0.05, 0.18, and 0.94, for J, H, and K. In
other words, in the Ks band R CrA could affect the Ks intensities while in the
J band it should have only little influence. However, the relation between
$I_{\rm K}$ and $A_{\rm V}^{\rm NICER}$ does not indicate any gradient in ISRF
across the field from East to West.

According to Fig.~\ref{fig:correlation_AV_scat}, observed Ks intensities are 
below the predicted curve when extinction exceeds 4--5$^{\rm m}$. The effect
may be a direct indication of the shadowing produced by nearby optically thick
regions. As was seen in Fig.~\ref{fig:analytic_colden}, the sightlines with
$A_{\rm V}>5^{\rm m}$ are already quite close to the filament.  The effect is
not visible in Fig.~\ref{fig:JHKfit} because there the corresponding curves
were fitted to surface brightness data without knowledge of an absolute
$A_{\rm V}$ scale. The effect could mean that column densities are 
underestimated for the corresponding lines-of-sight. However, this conclusion
is not supported by the correlation in Fig.~\ref{fig:ext_correlation} where,
at high extinctions, the NICER method tends to predict lower extinction
values. 

Because of the differences in the radiation field and its attenuation by the
dense parts of the cloud the same parameters that were successfully used to
translate surface brightness into column density on the northern side, may not
work equally well on the southern side. In Sect.~\ref{sect:comparison} only
column density estimates north of the filament were discussed. However,
Figs.~\ref{fig:analytic_colden} and \ref{fig:extinction_map} show that on the
southern side the surface brightness method has predicted significantly higher
column densities than the NICER method. This is just an indication that the
parameter values employed in the surface brightness method are not global
constants and that, in the case of optically very thick structures, changes
can take place over short distances. 
Isotropic radiation field and optical depths below $A_{\rm V}\sim 20^{\rm m}$
were quoted as pre-requisites for reliable conversion between surface
brightness and column density. Our field is clearly in contradiction of both
rules. Nevertheless, in the region where the necessary parameters could be
determined the method was found to give reliable estimates. For very optically
thick regions like the central part of the imaged filament, the simple
approach used in this paper is not sufficient. In those cases numerical
modelling of the surface brightness observations may be required to improve
the accuracy of the estimates.

\subsection{Comparison with millimetre emission data}

Some information on the column densities for the centre of the filament comes
from Chini et al. (\cite{Chini2003}) who observed the R Corona Australis cloud
in 1.2\,mm continuum with the SIMBA instrument. The signal-to-noise ratio is
not high enough so that much structure could be seen in the region now
observed in NIR. However, in Fig.~2 of Chini et al. a local emission maximum,
consisting of millimeter sources MMS2 and MMS3, does coincide with our region
$e$. It is doubtful whether MMS2 and MMS3 are truly separate clumps or whether
their apparent division is caused by noise. Chini et al. do not suggest that
they would represent protostellar condensations. The sources reside within a
larger emission peak that agrees with the position of region $e$ and confirms
the presence of a local column density maximum there. The intensity of the mm
emission is $\sim$160\,mJy per 28$\arcsec$ beam. Using the parameters
suggested by Chini et al., a mass absorption cross section
$\kappa=0.37$cm$^2$g$^{-1}$ and a gas-to-dust ratio of 150, this corresponds
to a beam averaged hydrogen column density of $N({\rm H}_2)$=3.8$\times
10^{22}$cm$^{-2}$. Using the standard relation given by Bohlin et al.
(\cite{Bohlin78}), this corresponds to a visual extinction of $A_{\rm
V}$=40$^{\rm m}$. This is at least consistent with the observed scattered
light, which indicates an extinction well above 20 magnitudes. As mentioned, a
more quantitative value (or a strict lower limit) to $A_{\rm V}$ will be
derived from the NIR surface brightness data in a future paper with the help
of radiative transfer modelling. Figure.~1 of Chini et al. further indicates
the presence of higher column densities some distance to the west, coinciding
with the region $f/g$. Therefore, although the NIR scattered light suffers
from significant saturation, it is still able to point out the most opaque
regions and give some information on their small scale structure.

In the SIMBA observations of Chini et al. the effective mapping rate was
$\sim$1.8\,min per square arc minute which resulted in an 1-$\sigma$ noise of
17\,mJy per beam. Observations of an area similar to our NIR mapping, $4
\arcmin \times 8 \arcmin$, took approximately one hour. Assuming a dust
temperature of 16\,K the SIMBA sensitivity corresponds to $A_{\rm V}\sim
3^{\rm m}$ at the resolution of 24$\arcsec$. Based on
Fig.~\ref{fig:ext_correlation} the internal scatter in our NIR data
corresponds at low extinctions to $A_{\rm V}\sim 0.5-1.0^{\rm m}$. The lower
value may be closer to the truth because, as discussed in
Sect.~\ref{sect:ext_comparison}, most of the scatter appears to come from the
colour excess method. Adjusting for the difference in resolution and
sensitivity, a SIMBA type instrument would need between 13 and 52 hours of
observing time to reach similar accuracy. In other words, with similar
observing times similar accuracy could be reached with both the NIR and mm
instruments.

Compared with SIMBA the new generation of sub-millimeter bolometers, LABOCA and
SCUBA-II among the first, will significantly increase the efficiency of
dust emission observations at long wavelengths. However, similar
improvement is taking place in NIR instrumentation. As an example, the VIRCAM
instrument on ESO's VISTA telescope has an instantaneous field-of-view that is
two orders of magnitude larger than that of the SOFI instrument. The NIR and
sub-mm/mm wavelength observations remain largely complementary, especially
regarding the range of $A_{\rm V}$ they cover. However, with sufficiently long
integrations, the NIR scattering can provide maps with significantly higher
resolution.

\section{Conclusions} \label{sect:conclusions}

We have made deep near-infrared observations of a cloud filament in the Corona
Australis molecular cloud. The goal was to detect NIR scattered light towards
the cloud and to confirm that this surface brightness can be used to derive
reliable column density maps. The column density estimates were made using
methods presented by Padoan et al. (\cite{Padoan2006}) and Juvela et a.
(\cite{Juvela2006}). The photometry of the background stars and the NICER
method were used to derive an independent extinction map.

Our results show that
\begin{itemize}
\item Surface brightness could be detected over most of the imaged fields
where, according to the NICER analysis, the $A_{\rm V}$ ranges from 
$1.8^{\rm m}$ to more than $30^{\rm m}$.
\item The NIR intensities and their $A_{\rm V}$ dependencies
are consistent with the surface brightness being caused by dust scattering. No
clear indication of additional dust emission at low $A_{\rm V}$ was detected.
\item In regions below $A_{\rm V}\sim 15^{\rm m}$ the column density estimates 
derived using scattered light and the colour excesses of background stars agree with each
other.
\item In our case the surface brightness data allow construction of a column
density maps with resolution a few times better than that provided by the
background stars.
\item In the interval $A_{\rm V}=15-20^{\rm m}$ the surface brightness data 
predicts up to 50\% higher extinction values than the colour excess method. We
interpret this as bias in the colour excess method caused by the strong
extinction gradients and the small number of background stars.
\item The absence of background stars prevents the NICER method from providing
reliable estimates for the extinction at the centre of the filament. The
morphology of surface brightness maps is still able to point out local
extinction maxima. However, quantitative estimates of $A_{\rm V}$ (or its
lower limits) will require modelling of the cloud and the radiation field.
\end{itemize}

The results show that near-infrared scattering can be used for mapping of
quiescent interstellar clouds. The changes in dust grain properties within
dense clouds and small scale variations in the gas-to-dust ratio remain
possible sources of uncertainty when any dust tracers are used to infer the
distribution of total cloud mass (e.g. Padoan et al. \cite{Padoan2006}). On
the other hand, by providing high resolution images of the dust distribution
the near-infrared scattering also provides a useful tool for the study of
these questions.

\acknowledgements

M.J. and V.-M.P.  acknowledge the support of the Academy of Finland Grants no.
206049, 115056, 107701, and 124620. P.P. was partially supported by the NASA
ATP grant NNG056601G and the NSF grant AST-0507768.

\appendix

\section{NIR correlations} \label{sect:NIR_correlations}

\begin{figure} 
\resizebox{\hsize}{!}{\includegraphics{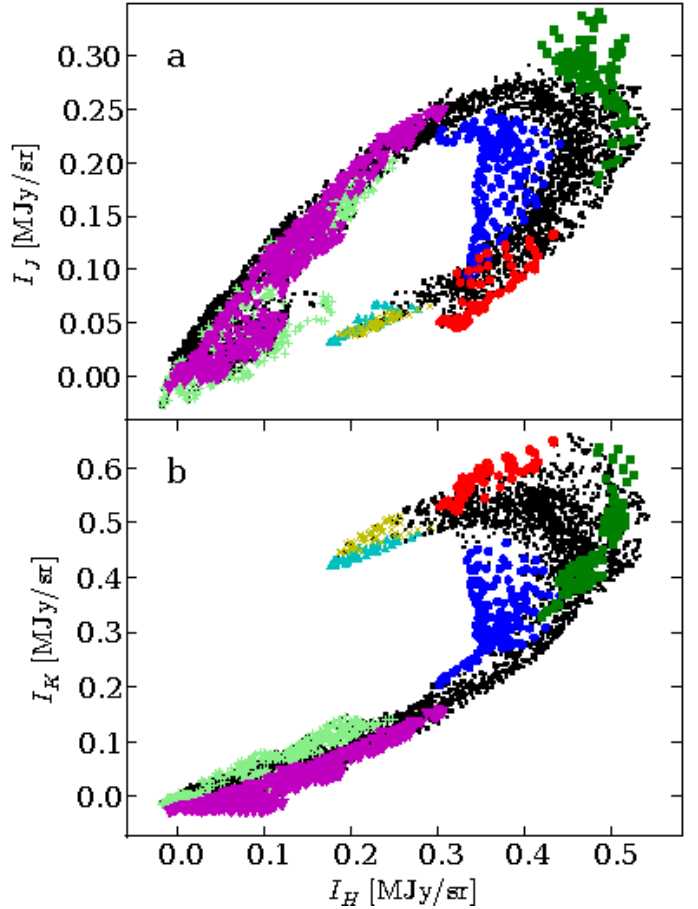}} 
\caption{
Correlations between the observed NIR surface brightness values. The plots
complement Fig.~\ref{fig:plot_correlation_map_JHK} that showed the remaining
correlations between the J and H bands ({\em frame a}) and H and K bands ({\em
frame b}).
} 
\label{fig:HJ_HK} 
\end{figure}

Figure~\ref{fig:HJ_HK} shows correlations between the observed J and H and the
H and K surface brightness values. The regions that were identified in 
Fig.~\ref{fig:plot_correlation_map_JHK} are also here plotted with different
symbols.  The regions $c$--$g$ correspond to distinct areas in all
correlations. This indicates that the spectra of scattered light are to some
extent anomalous in these regions (see Sect.~\ref{sect:qualitative}). On the
other hand, regions $a$ and $b$ are separated clearly only along the K--axis.
This suggest that they could be caused by observational effects, i.e., an
artificial gradient in the K band image. 

\begin{figure} 
\resizebox{\hsize}{!}{\includegraphics{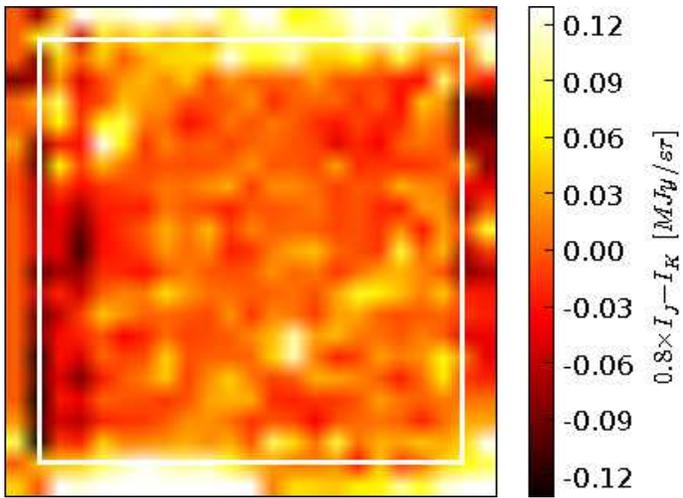}} 
\caption{
The colour 0.8$\times$J-K in the field OFF1 after stars have been removed
with median filtering. The image shows some systematic colour variations at
the image borders. The box indicates the extent of the image borders that in
our analysis was mask out.
} 
\label{fig:check_off1} 
\end{figure}

Similar effects are seen in the off fields where the expected signal from dust
should be negligible. Figure~\ref{fig:check_off1} shows for the field OFF1 the
colour 0.8$\times$J-K after stars have been removed using a median filter of
50$\times$50 pixels. At the image borders there are systematic colour
variations at a level of close to 0.1\,MJy\,sr$^{-1}$, i.e. at a level
comparable to the regions $a$ and $b$. 
This could be caused, e.g., by an imperfect illumination correction.
In our analysis the image borders were
masked out. However, border effects in individual frames could still explain
some of the features seen in Fig.~\ref{fig:plot_correlation_map_JHK} because
the jitter pattern was larger than the width of the rejected borders. In this
respect the analysis could still be improved by applying a mask already to
individual frames instead of the final image.

\end{document}